\documentclass[prl,twocolumn,showpacs,preprintnumbers,amsmath,amssymb,superscriptaddress]{revtex4}

\usepackage{graphicx}
\usepackage{dcolumn}
\usepackage{bm}

\begin{document}

\title{Ultra-sensitive Diamond Magnetometry Using Optimal Dynamic Decoupling}

\author{L. T. Hall}
 \email{lthall@physics.unimelb.edu.au}
\affiliation{Centre for Quantum Computing Technology, School of Physics, University of Melbourne, Victoria 3010, Australia}%
\author{C. D. Hill}
\affiliation{Centre for Quantum Computing Technology, School of Physics, University of Melbourne, Victoria 3010, Australia}%
\author{J. H. Cole}
\affiliation{Institute f\"ur Theoretische Festk\"orperphysik and DFG-Centre for Functional Nanostructures (CFN), Karlsruhe Institute f\"ur Technologie, 76128 Karlsruhe, Germany}
\author{L. C. L. Hollenberg}
\affiliation{Centre for Quantum Computing Technology, School of Physics, University of Melbourne, Victoria 3010, Australia}%


\begin{abstract}
New magnetometry techniques based on Nitrogen Vacancy (NV) defects in diamond have received much attention of late as a means to probe nanoscale magnetic environments. The sensitivity of a single NV magnetometer is primarily determined by the transverse spin relaxation time, $T_2$. Current approaches to improving the sensitivity employ crystals with a high NV density at the cost of spatial resolution, or extend $T_2$ via the manufacture of novel isotopically pure diamond crystals. We adopt a complementary approach, in which optimal dynamic decoupling techniques
extend coherence times out to the self-correlation time of the spin bath.
This suggests single spin, room temperature magnetometer sensitivities as low as 5\,pT\,Hz$^{-1/2}$ with current technology.
\end{abstract}


\maketitle
Room temperature single spin magnetometry using the nitrogen-vacancy (NV) centre in diamond has the potential to revolutionize nanoscale imaging  through fundamentally new detection modes \cite{Che04,Deg08,Tay08,Maz08,Bal08,Bal09,Col08,Hal09}.
Proposals to image nanoscale environments exhibiting static (DC) and oscillatory (AC) magnetic fields using NV systems \cite{Deg08,Tay08} have since been demonstrated experimentally \cite{Maz08,Bal08,Bal09}. By exploiting measured changes in quantum decoherence \cite{Col08,Hal09} these techniques have been extended to include more general classes of randomly fluctuating (FC) magnetic fields, with comparable sensitivity to the AC case \cite{Hal09}. One prominent reason such room temperature single spin detection techniques are of interest is because of the potential to develop fundamentally new imaging modes for biological systems with nanometer resolution \cite{Hal09-2}.

The sensitivity of an NV magnetometer is governed by the transverse spin relaxation (dephasing) time $T_2$, which for AC detection using isotopically-pure diamond has been demonstrated at 4\,nT\,$\mathrm{Hz}^{-1/2}$ \cite{Bal09}. Many important detection problems, particularly those related to biology \cite{Hal09-2}, stand to gain significant improvements with increased coherence times. In this paper we show how optimal dynamic decoupling techniques \cite{Uhr07} can be exploited to increase sensitivities by over two orders of magnitude. With sensitivities in the pT\,$\mathrm{Hz}^{-1/2}$ regime, and nanoscale spatial resolution, the ultra-sensitive NV magnetometer proposed here may have profound implications for nano-bio imaging and sensing.

\begin{figure}
\includegraphics[width=8.5cm]{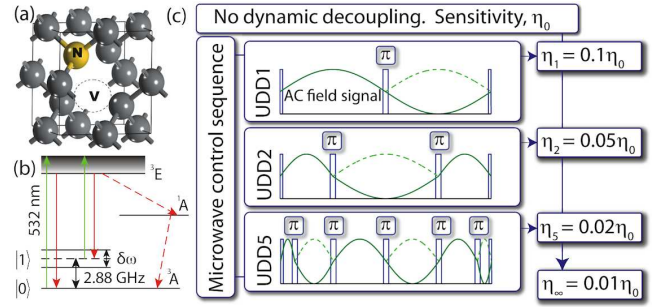}\\
\caption{(color online) (a) NV-centre diamond lattice defect. (b) NV spin
detection through optical excitation and emission cycle. Magnetic sublevels $m_s = 0$ and $m_s = \pm1$ are split by a $D$=2.88 GHz crystal field. Degeneracy between the $m_s = \pm1$ sublevels is lifted by a Zeeman shift, $\delta\omega$.
Application of 532 nm green light induces a spin-dependent photoluminescence and pumping into the $m_s = 0$ ground state.
(c) Examples of controlled-AC fields (solid) as seen by the NV centre (dashed) in the presence of the 1$^{\mathrm{st}}$, 2$^{\mathrm{nd}}$ and 5$^{\mathrm{th}}$ UDD sequence. Negative regions of the AC trace are mapped to positive, ensuring maximal phase accumulation of the NV spin. Slow FC fields, such as the surrounding nuclear spin bath, will be suppressed, permitting the detection of AC and fast FC fields with greater sensitivity.} \label{techschem}
\end{figure}

It is well known that coherence times may be improved with the use of concatenated (CDD)\cite{Wit07}, random and periodic dynamic decoupling schemes \cite{Wit07,Kho07}. The more recent Uhrig (UDD) scheme was shown to be optimal for decoupling a spin qubit from a bosonic bath\,\cite{Uhr07}, and has since been applied to a much broader class of systems in which dephasing is the dominant decoherence channel, including classical noise \cite{Uhr08} and spin bath systems\,\cite{Lee08}. This optimality stems from the fact that the required resources of the UDD sequence grows linearly with the order to which the environmental effects are suppressed\,\cite{Lee08}. However, UDD is unable to suppress longitudinal relaxation, whereas CDD can, but does so at exponential cost. The use of UDD is ideally suited to the NV centre owing to long relaxation times ($T_1>1$\,s \cite{Bal09}), even at room temperature. The large Debye temperature allows for negligible decoherence via phonon excitation of the crystal lattice, and a large zero field splitting (2.88 GHz) of the ground state magnetic sublevels prevents longitudinal spin-spin relaxation. Hence, for the NV centre, longitudinal relaxation may be neglected, ensuring UDD is the optimal decoupling method.

Both AC and FC magnetometry schemes are based upon a spin-echo microwave control sequence, in which a $\pi$ pulse is used to flip the qubit at the half way point of its evolution (Fig.\,\ref{techschem}(c)), suppressing any quasi-static effects of the spin bath. The AC scheme is concerned with detection of fields of the form $b_{\mathrm{ac}} = b_0 \sin(\nu t )$, where the $\pi$ pulse coincides with $t=\pi/\nu$, ensuring a non-zero integral of the field trace, and hence maximal phase shift of the NV spin. In this paper, we incorporate UDD into the NV magnetometry scheme and show that the sensitivity of a single NV centre magnetometer may be as low as 5 pT\,$\mathrm{Hz}^{-1/2}$ with the use of existing technology.

Recently it was shown that a particularly intuitive analysis of a spin qubit placed in a slowly fluctuating classical magnetic field could be performed by expanding the time dependent field as a Taylor series \cite{Hal09}. Whilst this may seem like a special case, this technique applies to a more general class of pure-dephasing quantum problems in which longitudinal relaxation may be ignored. We investigate the effect of UDD on a spin qubit placed in such a field, and show that it is the $n^{\mathrm{th}}$ UDD sequence that suppresses the effect of all terms up to and including order $n$ in the Taylor expansion of the field. These results are applied to the NV centre and used to obtain improved sensitivities for NV based magnetometry.

An NV centre aligned along the $z$ axis interacting with an external magnetic field is described by the Hamiltonian $\mathcal{H} = \mathcal{H}_\mathrm{zfs} + \mathcal{H}_\mathrm{ext} + \mathcal{H}_\mathrm{int}$. The first term describes the zero field splitting of the ground state Zeeman levels, $\mathcal{H}_\mathrm{zfs} = \hbar DS_z^2$, where $D=2.88$\,GHz. The interaction with an external magnetic field $\mathbf{B}_\mathrm{ext}(t)$ is described by $\mathcal{H}_\mathrm{ext}$. The fields we consider here are small relative to $D$ and are hence unable to induce a spin-flip, permitting us to ignore all $S_{x,y}$ terms, giving $\mathcal{H}_\mathrm{ext}=\gamma_{\mathrm{nv}}\mathbf{S}\cdot\mathbf{B}_\mathrm{ext}\approx\gamma_{\mathrm{nv}}B^{(z)}_\mathrm{ext}S_z$, where $\gamma$ is the electron gyromagnetic ratio. For simplicity, we put $B^{(z)}_\mathrm{ext}\equiv B_\mathrm{ext}$. The final term describes the interaction with the paramagnetic environment of the diamond crystal. As we will see, for resolving the non-unitary dynamics of the reduced density matrix of the NV centre, these interactions may be subsumed into a single `internal magnetic field,' $B_\mathrm{int}(t)$. We define the fluctuation regime of the external/internal environment via the dimensionless numbers, $\Theta_\mathrm{ext} = (\gamma\sigma_0^\mathrm{ext}\tau_\mathrm{ext})^{-1}$ and $\Theta_\mathrm{int} = (\gamma\sigma^\mathrm{int}_0\tau_\mathrm{int})^{-1}$, where $\sigma_0^\mathrm{ext}/\sigma_0^\mathrm{int}$ are the RMS field strengths, and $\tau_\mathrm{ext}/\tau_\mathrm{int}$ are the correlation times of the internal and external environments. Rapidly and slowly fluctuating fields satisfy $\Theta\gg1$ and $\Theta\ll1$ respectively.

An arbitrary time-dependent magnetic field may be decomposed as a Taylor series in $t$:
$B(t) = \sum_{k=0}^\infty a_kt^k$. The validity of this expansion rests upon the condition that $a_{n+1}t^{n+1}<a_nt^n\,\,\mathrm{for\,\,all}\,\, n$. As shown in \cite{Hal09} this is satisfied for $t<\tau_\mathrm{int}/\sqrt2$. For times $t\gg\tau_\mathrm{int}$ the qubit will exhibit motional-narrowing behaviour.

In many cases of practical interest, $B_\mathrm{int}$ is the sum of fields from a large number of dipoles. This implies that the $\{a_j\}$ are normally distributed, with zero mean at room temperature and variance $\sigma_j^2 = \left\langle a_j^2\right\rangle$. This leads to the following dephasing envelope, $\mathcal{D}(t) = \prod_{j=0}^\infty\exp\left[-\left(\Gamma_jt\right)^{2j+2}\right]$, where
\begin{eqnarray}
\Gamma_j&=&\left(\frac{1}{\sqrt2}\frac{\gamma\sigma_j}{j+1}\right)^{1/(j+1)}.\label{Hallexp}
\end{eqnarray}

Since $\Gamma_0\gg \Gamma_k\,\forall k\geq 1$, $\Gamma_0$ serves to define the free induction decay time, $T_2^* = 1/\Gamma_0$. For a 1.1\% $^{13}$C bath, the variance of the magnetic field is given by $ \sigma_0^2 = \sum_i\langle B_i^2\rangle$. Approximating this sum by an integral gives
$\sigma_0 \approx\sqrt{\frac{2\pi}{3}}\frac{\mu_0}{4\pi}n_\mathrm{c}g_\mathrm{c}\mu_\mathrm{N}\approx2\,\mu\mathrm{T}$,
and $T_2^*\approx4\,\mu$s, in good agreement with \cite{Tay08,Miz09}. Similarly, using an 0.3\% $^{13}$C bath yields $T_2^*\approx15\,\mu$s, in agreement with \cite{Bal09}.

The Hahn-echo sequence removes the effect of a static field on the system (though there are an infinite number of sequences that achieve this), as each $\pi$ pulse effectively sends $B\rightarrow-B$. For a field described by $\sum_k a_kt^k$, the Hahn-echo sequence will remove the effect of the $a_0$ term, and modify all other terms as $a_j\mapsto \left(1-2^{-j}\right)a_j$ \cite{Hal09}. For an NV centre, the correlation time of the environment is dictated by interactions between $^{13}$C nuclei.
A straight-forward calculation shows
  $\tau_\mathrm{int} \sim \sqrt{\frac{6}{\pi}}\frac{4\pi\hbar}{\mu_0}/(n_\mathrm{c}g_\mathrm{c}^2\mu_\mathrm{N}^2) \,=\,15\,\mathrm{ms}$.
Substituting this into Eq.\,(\ref{Hallexp}) for $j=1$ gives $\Gamma_1 = 2.1$ kHz. We identify $T_2 = 1/\Gamma_1 = 400$ $\mu$s, again in agreement with experiment \cite{Tay08,Miz09}. For an 0.3\% $^{13}$C bath we achieve $T_2 = 1.5$\,ms as seen in \cite{Bal09}. We do not define $\tau_\mathrm{int}$ via interaction between $^{13}$C nuclei and any background fields, $B_0$, since this manifests as decays and revivals on timescales of $\tau_\mathrm{r}\sim\gamma_\mathrm{c}B_0$ and does not represent a true loss of information.

We may wish to modify our $\frac{\tau}{2}-\pi-\frac{\tau}{2}$ pulse sequence in order to remove the effect of the $a_1t$ term.
If we apply pulses at $t=\frac{\tau}{4}$ and $t=\frac{3\tau}{4}$, we find that the effects of both $a_0$ and $a_1$ terms are suppressed. In general, suppression of all terms up to and including order $n$ will require $n+1$ $\pi$-pulses. We define $\tau_{n,k}$ as the time at which the $k^\mathrm{th}$ pulse is applied in the sequence that suppresses all field components up to, and including, order $n$. Determination of the $n+1$ elements of the set $\mathcal{P}_n=\bigl\{\tau_{n,0}\,,\ldots,\,\tau_{n,n+1}\bigr\}$ will require the solution of the following set of $n+1$ algebraic equations for $\tau_{n,k}$. For $a_0,\,2\sum_{k=1}^{n+1} \left(-1\right)^{k-1}\tau_{n,k} + \left(-1\right)^{n+1}\tau = 0$; for
$a_1,\,2\sum_{k=1}^{n+1} \left(-1\right)^{k-1}\tau_{n,k}^{2} + \left(-1\right)^{n+1}\tau^{2} = 0$; up to
\begin{eqnarray}
  a_n&:&\,\,\,2\sum_{k=1}^{n+1} \left(-1\right)^{k-1}\tau_{n,k}^{n+1} + \left(-1\right)^{n+1}\tau^{n+1} = 0,\label{con}
\end{eqnarray} 
To this point, there is some freedom in our choice of pulse sequence, as Eqs.\,(\ref{con}) are satisfied to $n=1$ by both CDD and UDD. However, if we solve Eqs.\,(\ref{con}) to $n=2$, we find $\mathcal{P}_2 = \left\{\frac{\tau}{2}\left(2-\sqrt2\right),\,\frac{\tau}{2},\,\frac{\tau}{2}\left(2+\sqrt2\right)\right\}$, which is the third UDD sequence. It is reasonable to conjecture that the $n^\mathrm{th}$ component of the field, $a_nt^n$, will be suppressed by the $(n+1)^\mathrm{th}$ UDD sequence, which we prove below.

For an interrogation time of $\tau$, the time of application of the $k$th pulse in the $n$th UDD sequence is given by
  $\tau_{n,k} = \tau\sin^2\left(\frac{\pi k}{2n+2}\right)$ \,\cite{Uhr07},
where $1\leq k\leq n$. 
We wish to show that the $(n+1)^\mathrm{th}$ UDD sequence suppresses the effect of terms up to and including the $n\mathrm{th}$ term in the Taylor expansion of a time dependent magnetic field; and that the effect of all terms beyond $n$ will be reduced.

The phase accumulation of a spin qubit is proportional to the time integral of the magnetic field to which it is exposed. Each $\pi$ pulse exchanges the basis states of the qubit Hilbert space, which has the same effect on their relative phase as mapping $B\mapsto-B$. Recall from Eq.\,(\ref{con}) that 
the action of the $(n+1)^\mathrm{th}$ Uhrig pulse sequence will be to modify each of the $a_j$ via
  $a_j \mapsto \left[2\sum_{k=1}^{n+1}(-1)^{k-1}\sin^{2j+2}\left(\frac{\pi k}{2n+4}\right) + (-1)^{n+1}\right]a_j$.
To prove our claim, we must firstly show that,
\begin{eqnarray}
  2\sum_{k=1}^{n+1} \left(-1\right)^{k-1}\sin^{2m}\left(\frac{\pi k}{2n+4}\right) &=& \left(-1\right)^{n} \label{lem1} .
\end{eqnarray}
Using $N\equiv n+2$, $\sin(x) = \frac{1}{2i}\left(e^{ix}-e^{-ix}\right)$ and expanding as a binomial series in $j$, we arrive at
\begin{eqnarray}
  \mathrm{LHS} &=& 2^{1-2m}\sum_{j=0}^{2m}(-1)^{j+m}\left(
                                                           \begin{array}{c}
                                                             2m \\
                                                             j \\
                                                           \end{array}
                                                         \right)\sum_{k=0}^{n-1}e^{i\pi a_{jmN}k},\label{corproof1}
\end{eqnarray}where $a_{jmN}\, =\, \frac{m-j-N}{N}$. Note that we have added the $k=0$ term, since $\sum_{j=0}^{2m}\left(^{2m}_{\,\,j}\right)(-1)^j=0$. Since we have restricted ourselves to $m\leq N-1$, we have that $e^{i\pi a_{jmN}}\neq1$, so we are free to sum over $k$ as a geometric series in Eq.\,(\ref{corproof1}), which is not possible for $m\geq N$. This gives
  $\mathrm{LHS}=2^{-2m}(-1)^{2m+N}\sum_{j=0}^{2m}\left(^{2m}_{\,j}
                                                         \right)$,
which is just the sum of terms in the ($2m+1^\mathrm{th}$) row of Pascal's triangle, and evaluates to $2^{2m}$, proving Eq.\,(\ref{lem1}).

Replacing $N$ with $n+2$, we can see that $a_j\mapsto0\,,\,\forall\,j\leq n$. Hence all terms in the Taylor expansion of $B(t)$ up to and including order $n$ are zero. Furthermore, since $\sin^2(x)\leq1$, the term in the square brackets is always less than 1, hence the effect of the remaining $a_j$ is reduced.

Two immediate consequences are that the $(n+1)^\mathrm{th}$ UDD sequence will suppress dephasing to $n^\mathrm{th}$ order; and that dephasing effects beyond $n^\mathrm{th}$ order will be reduced. If $a_0,\,\ldots,\,a_n = 0$, then $\left\langle a_0^2\right\rangle,\,\ldots,\,\left\langle a_n^2\right\rangle = 0$. By Eq.\,(\ref{Hallexp}) all rates, $\Gamma_0,\,\ldots,\,\Gamma_n$, are zero. In what follows, we examine how these ideas may be implemented to improve the sensitivity of a qubit magnetometer.

\begin{figure}
\begin{center}
  \includegraphics[width=8.5cm]{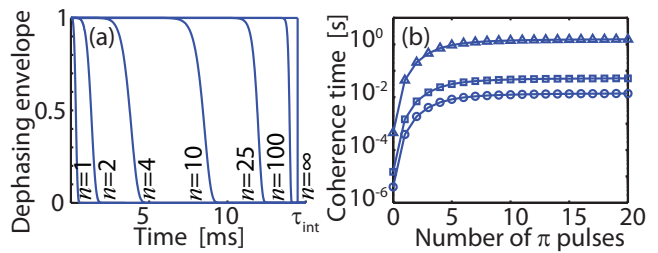}\\
  \caption{(colour online)(a) Dephasing envelopes, $\mathcal{D}^{(n)}$, for an NV centre in a 1.1\% $^{13}$C bath for $n\,\,\pi$ pulses. As $n\rightarrow\infty$, $\mathcal{D}^{(n)}$ approaches the Heaviside step function, $H(\tau_\mathrm{ext}-t)$. (b) Effect of the number of $\pi$ pulses on NV coherence times for different $^{13}$C concentrations. In each case, the coherence time is limited by $\tau_\mathrm{int}$.}\label{dectimes}
\end{center}
\end{figure}

From the above analysis, we see that the application of UDD sequence of any order will decrease the NV dephasing rate due to the internal fluctuating magnetic field. This allows us to extend the interrogation time, and hence improve the sensitivity to an external magnetic field, $B_\mathrm{ext}(t)$. Clearly the dynamics of $B_\mathrm{ext}$ will be an important factor in ensuring that the effect of the external field is not also suppressed by the pulse sequence. Simple examples include telegraph signals switching in sync with the UDD sequence, or an AC field of controllable frequency whose nodes coincide with each $\pi$ pulse, which could be realised by a single spin or ensemble of spins being driven by a controllable microwave field. Sensitivity to rapidly fluctuating fields will also be improved, since fields with correlation times shorter than the interrogation time will not be refocussed by the UDD sequence.

We denote the dephasing envelope in the presence of the $(n+1)$th pulse sequence as $\mathcal{D}^{(n)}(t) = \prod_{k=n}^\infty\mathcal{D}^{(n)}_k(t)$ [Fig.\,\ref{dectimes}(a)].
In the presence of background dephasing described by Eq.\,(\ref{Hallexp}), the minimum induced phase from $B_\mathrm{ext}(t)$ that may be measured is $\Delta \phi(b)= [C\sqrt N\mathcal{D}^{(n)}(\tau)]^{-1}$, where $C$ describes photon shot noise and imperfect collection \cite{Tay08}, and $N$ is the number of measurements taken. Typically $C<0.3$, however vast improvements have recently been demonstrated by entangling the NV spin with proximate nuclear spins, permitting repetitive readout of the NV spin state \cite{Ste09,Jia09}.


We now discuss the relevant detection protocols and associated sensitivities for different fields to which these techniques apply. For a telegraph signal (square waveform) switching between $\pm B_0$ in sync with each $\pi$ pulse, the qubit will acquire the maximum possible phase for a given interrogation time, $\Delta\phi = \gamma B_0\tau$. This gives a magnetic field sensitivity of $\eta^{(n)}_\mathrm{ts} = B_0\sqrt T = \left[C\gamma\sqrt\tau\mathcal{D}^{(n)}(\tau)\right]^{-1}$.
For all cases where $\Theta_\mathrm{int}\ll1$, we have that $\Gamma_k\gg\Gamma_{k+1}$, and $\Gamma_k>\Gamma^{(n)}_{k}$\,\cite{Hal09}, so we may approximate the total dephasing envelope, $\mathcal{D}^{(n)}$, by the its leading order contribution. That is $\mathcal{D}^{(n)}(t)   \sim\exp\left[-\left(\Gamma_{n+1}t\right)^{2n+4}\right]$,
implying the optimal interrogation time is $\tau = \Gamma^{-1}_{n+1}\left(4n+8\right)^{1/(2n+4)}$.
We then find the sensitivity to be bounded above by $\eta_+^{(n)}<\frac{1}{C\gamma}\sqrt{f_e\Theta_\mathrm{int}^{-1/n}}$.
This upper bound, together with the actual sensitivity (see below), is plotted in Fig.\,\ref{sens}(a) for an NV.
Notice that, as $\Theta_\mathrm{int}\rightarrow1$, there is little to be gained by applying UDD.

We now compute the sensitivity of the probe by taking into account the effect of the $n^\mathrm{th}$ order pulse sequence on the $n^\mathrm{th}$ order Taylor coefficient. This results in improved sensitivity beyond that indicated above, as a reduction in the $a_k$ leads to a reduction in the total decoherence rate, and hence an extended interrogation time.

\begin{figure}
\begin{center}
  \includegraphics[width=8.5cm]{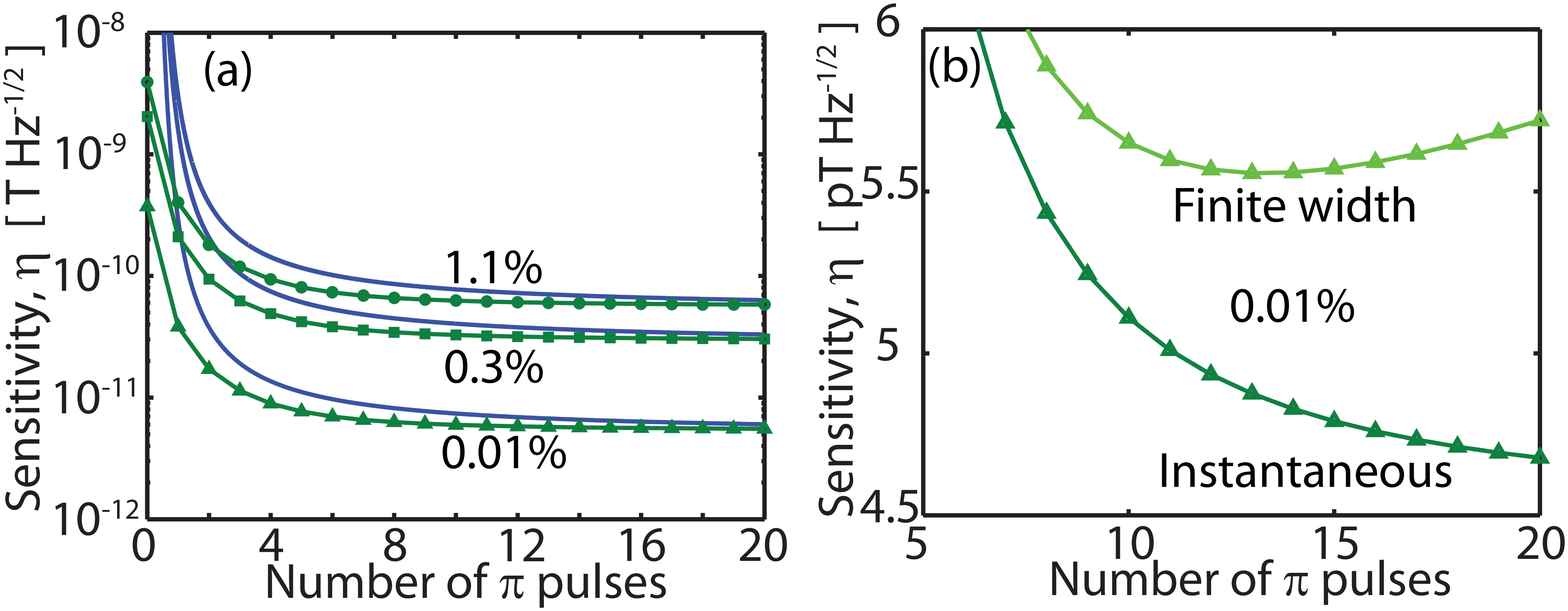}\\
  \caption{(colour online)(a) Fundamental shot-noise sensitivity limits assuming $C=1$ and a synchronised square wave-form for different $^{13}$C concentrations. Sensitivities to controlled AC signals are modified by a factor of $\pi/2$; those due to FC signals satisfying $\Theta_\mathrm{ext}>1$ are modified by a factor of $2\Theta_\mathrm{ext}$. Smooth curves show the analytic upper bound on the sensitivity, $\eta_+^{(n)}$. (b) Effect of finite pulse widths of 50\,ns on sensitivities. The optimal sensitivity occurs for $n=13$ $\pi$ pulses.}\label{sens}
\end{center}
\end{figure}

The dephasing rates are found via
Eq.\,(\ref{Hallexp}),
where
  $\sigma^{(n)}_j \mapsto \left[2\sum_{k=1}^{n+1}(-1)^{k-1}\sin^{2j+2}\left(\frac{\pi k}{2n+4}\right) + (-1)^{n+1}\right]\sigma_j. $ 
The actual dephasing time due to the combined effect of all the $\Gamma_k^{(n)}$ will be given by the solution to $\sum_{k=n+1}^\infty \left(\Gamma_k^{(n)}t\right)^{2k+2}-1 = 0$ for $t$, and are plotted against the number of pulses used in the sequence in Fig.\,\ref{dectimes}(a). From this we see that dephasing times asymptote to the correlation time of the bath. This is to be expected, since information lost to the bath cannot be recovered by control of the NV spin alone.
Retaining all terms, the sensitivity is then
$\eta^{(n)}(\tau) = \frac{1}{\gamma\sqrt\tau}\exp\left[\sum_{k=n+1}^\infty \left(\Gamma_k^{(n)}\tau\right)^{2k+2}\right]$.
By minimising this function with respect to $\tau$ we determine the optimal sensitivity,
$\eta^{(n)}_\mathrm{ts}$, as shown in Fig.\,\ref{sens}(a).

In many proposals \cite{Deg08,Tay08} magnetic resonance techniques are used to drive the sample magnetisation at some controlled frequency. By synchronising frequencies with the chosen pulse sequence, a piecewise continuous sinusoidal signal may be produced [Fig.\,\ref{techschem}(c)]. The sensitivity is hence modified by a factor of $\pi/2$, $\eta_\mathrm{ac}(\tau) = \frac{\pi}{2}\eta_\mathrm{ts}^{(n)}$.

Instead of sensing a controlled signal, we may wish to characterise a random, rapidly fluctuating ($\Theta_\mathrm{ext} \gg1$) external field, $B_\mathrm{ext}$. Rather than yielding an identical phase shift in each run, such a field will introduce an additional dephasing component and may be detected as a perturbation in the dephasing rate \cite{Hal09}. The dephasing envelope will be modified by a factor of $\mathcal{D}_\mathrm{ext}\left(\tau\right) = \exp\left(-\Gamma_\mathrm{ext}t\right)$, where $\Gamma_\mathrm{ext}= \frac{1}{2}\gamma_\mathrm{p}^2\sigma^2_\mathrm{ext}\tau_\mathrm{ext}$. The sensitivity with which $\sigma_\mathrm{ext}=\sqrt{\left\langle B_\mathrm{ext}^2\right\rangle-\left\langle B_\mathrm{ext}\right\rangle^2}$ may be measured is then
  $\eta_\mathrm{ffl}(\tau) = 2\Theta_\mathrm{ext}\eta_\mathrm{ts}^{(n)}$\cite{Hal09}, making the field more difficult to detect as the fluctuation rate increases. This is consistent with motional narrowing phenomena in NMR, in which high frequency noise is known to have a reduced effect on the sample $T_2$ as compared with quasi-static noise.

Coherent manipulation of the NV is achieved via a Rabi cycle, thus instantaneous $\pi$ pulses cannot be achieved in practice, and lead to additional decoherence effects. For a Rabi frequency of $\Omega$, the decoherence envelope is given by $\mathcal{D}_\mathrm{R} = [1 + (\gamma^2\sigma_0^2 t/\Omega)^2]^{-1/4}$ \cite{Dob09}, and typical pulse errors are $\approx$ 1\% \cite{Tay08}. Hence for $n$ $\pi$ pulses, the sensitivity will be worsened by a factor of $\Delta\eta \sim 0.99^{n+1}[1 + \frac{n+1}{4}(\sqrt\pi\gamma\sigma_0 /\Omega)^4]$, as shown in Fig.\,\ref{sens}(b). For a pulse width of 50\,ns, 13 $\pi$ pulses is found to be optimal, with $\eta_\mathrm{ts}^{(13)}\approx5.5$\,pT\,Hz$^{-1/2}$.

We have theoretically investigated the improvements associated with the application of the optimal Uhrig control sequence to an NV-based magnetometer. Results show that dephasing times due to the paramagnetic impurities are ultimately limited by the self correlation time of the fluctuating environment, thus NV magnetometer interrogation times may be extended by nearly four orders of magnitude beyond the free-induction decay time. In light of these results, we have shown that incorporation of UDD into current single NV magnetometer protocols may yield sensitivities below 5\,pT\,Hz$^{-1/2}$ at room temperature in the near future. Such techniques have the potential to yield great improvements to nano-scale sensing, particularly nano-biological processes occurring at room temperature.

\begin{acknowledgments}
This work was supported by the Australian Research Council.
\end{acknowledgments}

\bibliography{pulsebib}

\end{document}